\documentclass[twocolumn,english,aps,reprint,groupedaddress,notitlepage,nobibnotes,nofootinbib,preprintnumbers,showpacs]{revtex4-1}
\usepackage{graphicx,color,colortbl,amsmath,multirow,comment,slashed}
\usepackage[yyyymmdd,hhmmss]{datetime}
\usepackage[bookmarks=false]{hyperref}

\newif\ifdraft
\drafttrue
\newif\ifpreprint
\preprinttrue

\def\Section#1{\noindent\textsl{#1.\/}} 

\usepackage{color}

\def\fig#1{Fig.~{\ref{#1}}}

\def\spa#1.#2{\left\langle#1\,#2\right\rangle}
\def\spb#1.#2{\left[#1\,#2\right]}

\def\Ord{\mathcal{O}}

\def\eqn#1{Eq.~(\ref{#1})}

\def\eqns#1#2{Eqs.~(\ref{#1}) and~(\ref{#2})}

\def\NeqEight{{{\cal N}=8}}

\def\eps{\epsilon}
\def\nn{\nonumber}
\def\Section#1{\vskip .2 cm 
\noindent{\em #1:}}

\def\spa#1.#2{\left\langle#1\,#2\right\rangle}
\def\spb#1.#2{\left[#1\,#2\right]}

\def\eps{\epsilon}

\begin{document}

{\small \hbox{\hskip .5 cm UCLA/TEP/2020/101}  \hskip 4.7cm   {FR-PHENO-2020-001}  }

\title{Universality in the classical limit of massless gravitational scattering}

\author{Zvi~Bern${}^{a}$, Harald~Ita${}^{b}$,  Julio~Parra-Martinez${}^{a}$, 
Michael~S.~Ruf${}^{\,b}$}
\affiliation{
  $\null$\\
${}^a$Mani L. Bhaumik Institute for Theoretical Physics,\\
UCLA Department of Physics and Astronomy,\\
Los Angeles, CA 90095, USA\\ \vspace{-0.2cm}
\\ 
${}^b$Physikalisches Institut, Albert-Ludwigs-Universit\"at Freiburg,
 Hermann-Herder-Strasse 3, 79104 Freiburg, Germany
\\ \vspace{-0.2cm}
}

\begin{abstract}
We demonstrate the universality of the gravitational classical
deflection angle of massless particles through $\Ord(G^3)$ by studying
the high-energy limit of full two-loop four-graviton scattering
amplitudes in pure Einstein gravity as well as $\mathcal{N}\geq 4$
supergravity.  As a by-product, our first-principles calculation
provides a direct confirmation of the massless deflection angle in
Einstein gravity proposed long ago by Amati, Ciafaloni and Veneziano,
and is inconsistent with a recently proposed alternative.
\end{abstract}


\maketitle

\Section{Introduction}
The high-energy behavior of gravitational-scattering processes has a
long and interesting history as a fundamental probe of gravitational
theories at the classical and quantum level (see
e.g. Refs.~\cite{HighEnergyClassical, HighEnergyQuantum, ACV}).  The
simplicity of scattering in the high-energy limit makes it a natural
forum to extract information about high-orders in perturbation
theory. Indeed, using insight from string amplitudes and the
analyticity of scattering amplitudes, Amati, Ciafaloni and Veneziano
(ACV)~\cite{ACV} worked out the high-energy limit of massless graviton
scattering through $O(G^3)$ long before it became technically feasible
to compute two-loop scattering amplitudes in quantum field theory directly.
Using this they calculated the corresponding correction to the
gravitational deflection angle of massless particles in General
Relativity.

Recently the subject of scattering processes in gravitational theories has been
reinvigorated by the spectacular observation of gravitational waves by the
LIGO/Virgo Collaboration~\cite{LIGO}.  While scattering processes may seem
rather different from the bound-state problem of gravitational-wave generation,
the underlying physics is the same.  In particular, classical two-body
potentials can be extracted from scattering
processes~\cite{ScatteringToPotential}, including new state-of-the-art
calculations~\cite{3PMPRL, 3PMLong}. This approach leverages the huge advances
in computing quantum scattering amplitudes that stem from the modern unitarity
method~\cite{GeneralizedUnitarity} and from double-copy
relations~\cite{DoubleCopy} between gauge and gravity theories.

The possibility of using quantum scattering amplitudes to obtain the
classical deflection angle was also promoted by
Damour~\cite{DamourHighEnergy}, who used the ACV result for the
conservative scattering angle to impose constraints on classical
two-body Hamiltonians of the type used for gravitational-wave template
construction~\cite{EOB}.  In a very recent paper~\cite{DamourRecent},
however, Damour has cast doubt on the program of using quantum
scattering amplitudes to extract information on classical
dynamics. His central claim is that both the classical scattering
angle derived by ACV and the $\mathcal O(G^3)$ two-body Hamiltonian
derived in Ref.~\cite{3PMPRL,3PMLong} are not correct. His claims,
based on information obtained from the self-force (small mass ratio)
expansion~\cite{SelfForce} of the bound-state dynamics as well as
structural properties in the results of Ref.~\cite{3PMLong},
ultimately follow from the desire to have smooth transition between
massive and massless classical scattering.

In this Letter we confirm that the conservative scattering angle as
determined by ACV~\cite{ACV} is indeed correct.  Our confirmation
follows as a by-product of studying universality of the classical
scattering angle in massless theories. Remarkably, we find that the
scattering angle is independent of the matter content for a variety of
theories, implying graviton dominance in the high-energy limit.
Ref.~\cite{Bellini:1992eb} revealed hints of such dominance,
well-known at leading order~\cite{HighEnergyQuantum}, through analysis of
gravitino contributions.

Our study relies on having on hand the explicit expressions for massless
two-loop four-point amplitudes for $\mathcal N\geq 4$ supergravity
~\cite{BDDPR, N8FiniteRemainder,BoucherVeronneau:2011qv} and pure Einstein
gravity~\cite{FreiburgN0Ampl}.  
The latter result makes use of the latest advances in evaluating
multiloop amplitudes based on numerical unitarity followed by
analytic reconstructions~\cite{TwoLoopNumericalUnitarity}.  
Armed with the fully-evaluated amplitudes we then follow the
standard~\cite{EikonalReview} and widely used (see
e.g. Refs.~\cite{EikonalRecent, Melville:2013qca, Ciafaloni:2014esa, KoemansCollado:2019ggb, DiVecchia:2019kta}) extractions of the
scattering angle, using both impact parameter space and partial-wave
analyses.

For the case of $\mathcal N = 8$ supergravity a recent
paper~\cite{DiVecchia:2019kta} analyzes the eikonal phase through
$\Ord(G^4)$ using the two- and three-loop amplitudes from
Refs.~\cite{BoucherVeronneau:2011qv, HennThreeLoopN8}. The same
work~\cite{DiVecchia:2019kta} observes that the $\mathcal N = 8$
scattering angle matches the angle found by ACV through
$\Ord(G^3)$~\cite{ACV}, despite having different matter
content. Indeed, as we show here, this is not an accident, but part of
a general pattern.  Our explicit calculations for the $\Ord(G^3)$
contributions to the classical scattering angle in $\mathcal N \geq 4$
and pure gravity give the identical result as the angle found by ACV,
demonstrating its universality.

\Section{The classical limit of the amplitude}
We are interested in extracting the contributions to the conservative
classical scattering angle from the two-loop four-point scattering
amplitudes of Refs.~\cite{BDDPR,N8FiniteRemainder,
  BoucherVeronneau:2011qv, BoucherVeronneau:2011qv, FreiburgN0Ampl}.
Four-point scattering amplitudes depend on the kinematic invariants
$s$ and $t = -q^2$, which in the center of mass frame correspond to
the squared total energy and squared four-momentum transfer,
respectively. We consider the amplitude in the physical region $s>0,\,
t<0,\, u=-s-t<0$ (using a 
mostly-minus sign convention for the metric), commonly known as the $s$-channel.
The contributions in the amplitude relevant for the classical angle
corresponds to the large angular momentum limit, which for massless
particles is $J\sim \sqrt{s}\,b \gg 1$, where $b$ denotes the usual
impact parameter. In the absence of any other
kinematic scales such as masses in the momentum-space scattering
amplitude, the classical limit is equivalent to the Regge or
high-energy small-angle limit, $s/q^2\gg1$. It is straightforward to
argue that the singularity structure of massless scattering amplitudes
implies that only even loop orders can give rise to classical
contributions (see e.g.~Refs.~\cite{ACV, DiVecchia:2019kta} for a detailed
argument).  At one loop, in particular, this is directly tied to the
fact that no term behaves as $1/q$ which would be required to
contribute to the classical deflection angle.

Following Ref.~\cite{ACV}, we consider external graviton
states.  For simplicity we focus on the configuration where the
incoming and outgoing gravitons in the $s$-channel have identical helicity;
the situation where the incoming and outgoing gravitons have opposite helicity
gives the same final classical scattering angle.
We extract the classical scattering angle from the Regge limit of the
renormalized scattering amplitudes, which take the following
form, 
\begin{align}
  \mathcal{M}^{(0)}(s,q^2) &=\!\! \raisebox{-17pt}{\includegraphics[scale=0.5]{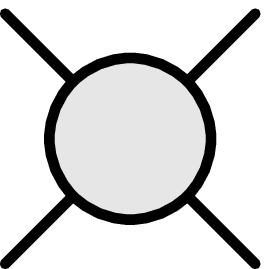}} \!\!= \mathcal{K} 8\pi G  s\biggl[\frac{s}{q^2} + 1 \biggr]  \,, 
\nn \\[5pt]
  \mathcal{M}^{(1)}(s,q^2)&= \!\!\raisebox{-17pt}{\includegraphics[scale=0.5]{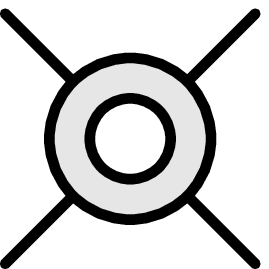}} \!\! 
 = 4 \mathcal{K} \,G^2 s^2 r_{\Gamma}\left(\frac{\bar\mu^2}{q^2}\right)^\epsilon \biggl[-\frac{2\pi i}{\epsilon} \frac{s}{q^2} 
\nn \\ 
  & \hskip 0.5 cm  \null  
+ \frac{1}{\epsilon} (2L+2 - 2\pi i ) 
+ F^{(1)}\biggr] \,,  \label{eq:TwoloopGenericAmpl}
\\[5pt]
\mathcal{M}^{(2)}(s,q^2) &=\!\!\raisebox{-17pt}{\includegraphics[scale=0.5]{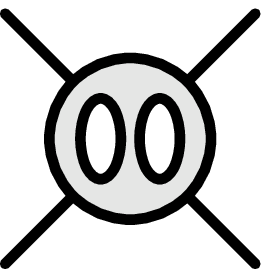}}\!\! = 2 \mathcal{K} \, G^3 s^3 \frac{r_{\Gamma}^2}{\pi} \left(\frac{\bar\mu^2}{q^2}\right)^{2\epsilon}  \biggl[ -\frac{2 \pi^2}{\epsilon^2}  \frac{s}{q^2}
 \nn \\
 & \hskip 0.55 cm  \null
- \frac{2\pi i}{\epsilon^2} (2L+2 - i\pi )
 - \frac{2\pi i}{\epsilon} F^{(1)} + F^{(2)} \biggr] \,, \nn 
\end{align}
where we dropped subdominant terms of
$\Ord(q^2/s)$ in the loop amplitudes, and
where $\mathcal{K}$ is a local factor depending on the external
states, $\bar\mu^2 \equiv 4\pi e^{-\gamma_E} \mu^2$ is a rescaled renormalization scale and $ r_{\Gamma}
\equiv e^{\epsilon\gamma_E} {\Gamma(1+\epsilon)\Gamma(1-\epsilon)^2}
/{\Gamma(1-2\epsilon)}$.
For convenience we introduced $L=\log(s/q^2)$, and the
finite remainders $F^{(i)}$, which depend on the theory and are implicitly defined in Eq.~\eqref{eq:TwoloopGenericAmpl}. This result is given in the conventional
dimensional regularization scheme, where all internal states and
momenta are analytically continued into $D=4-2\eps$ dimensions.  
For the purposes of this paper we only need  $ F^{(1)}$ to $\mathcal{O}(\epsilon)$ and $ F^{(2)}$ to $\mathcal{O}(\epsilon^0)$.
The two-loop infrared singular part is related
to the square of the one-loop amplitude via $[\mathcal{M}^{(1)}]^2/2
\mathcal{M}^{(0)}$ which follows from the fact that to all loop orders
the infrared singularity is given by an exponential of the ratio of
the one-loop and tree amplitudes~\cite{GravityIR,N8FiniteRemainder}.

The pure gravity one-loop amplitudes were originally computed in
Ref.~\cite{DunbarNorridge}. These were recomputed in an intermediate
step~\cite{InternalNotesGravity} of the two-loop analysis of
Ref.~\cite{TwoLoopGravityUV}.  This is matched by the expressions in
Ref.~\cite{FreiburgN0Ampl} that include the $\Ord(\epsilon)$
contributions.  The latter contributions are needed when
extracting the two-loop finite remainders in the presence of infrared
singularities, with the result,
\begin{align}
  F^{(1)}_{\rm GR} &= 2L^2 + 2i\pi L + 24\zeta_2  -\frac{87}{10}L  + \frac{841}{90} 
  \nn\\
		& \hskip 0.3 cm \null + \epsilon  \biggl[-\frac{2}{3} L^3
    -6 \zeta_2 L + 6 \zeta_3 + \frac{47}{20} L^2 -18 \zeta_2 -\frac{6913}{225} L   \nn\\
    & \hskip 0.3 cm \null +\frac{35597}{1200} 
		+ i \pi\Bigl(- L^2 + 2 \zeta_2 +10  L  + \frac{1957}{360} \Bigr)\biggr] \,,
\label{GRFiniteRemainder}
\end{align}
where $F^{(1)}_{\rm GR}$ is the pure gravity result for $F^{(1)}$ in
\eqn{eq:TwoloopGenericAmpl}. The
$\mathcal N \ge 4$ supergravity amplitudes can be found in
Ref.~\cite{DunbarNorridge, Bern:2011rj, BoucherVeronneau:2011qv} in a
scheme that preserves supersymmetry. For these cases, the Regge limit
of the $\Ord(\eps^0)$ contributions to the finite remainders can be
read off from Eq.~(4.6) of Ref.~\cite{Melville:2013qca}.

Ref.~\cite{FreiburgN0Ampl} provides the complete Einstein-gravity amplitude
needed for our analysis, including subdivergence
subtractions~\cite{tHooftVeltman,TwoLoopGravityUV,InternalNotesGravity}.
We note that these results pass highly nontrivial checks. The amplitude yields
the expected IR pole structure~\cite{GravityIR} and the net ultraviolet poles
cancel against the known counterterms~\cite{GoroffSagnotti, TwoLoopGravityUV}.
Furthermore the amplitude only has the poles in the Mandelstam variables $s$,
$t$ and $u$ dictated by factorization.  The amplitudes have also been validated
against results in the literature and independent computations. While not
directly relevant for the classical scattering angle, the results of
Ref.~\cite{FreiburgN0Ampl} also match the previously
computed~\cite{InternalNotesGravity} identical-helicity amplitude (in an all
outgoing momentum convention), corresponding to the case that
both incoming gravitons flip helicity.

Starting from the full four-graviton two-loop amplitude in pure Einstein
gravity~\cite{FreiburgN0Ampl}, we
extract the finite remainder in the Regge limit giving the result,
\begin{align}
 F^{(2)}_{\rm GR}=&-2\pi^2L^2 +4\pi^2L  -\frac{\pi^4}{90}+\frac{13403\pi^2}{675}-\frac{13049}{2160}
\nn \\
&\null 
+i\pi\biggl[\frac{4}{3}L^3 - \frac{47}{10}L^2
+\frac{5893}{150} L - 20\zeta_3 \nn \\
& \hskip 1.5 cm \null
 + \frac{2621\pi^2}{210} - \frac{106289}{3375}\biggr] \,.
\label{FiniteTwoLoop}
\end{align}
  The constant parts are scheme
dependent and in any case they do not affect the scattering angle.  A detailed
discussion of scheme dependence and its effects on the final angle, in the
context of IR regulators in $\NeqEight$ supergravity is found in Section 6 of
Ref.~\cite{DiVecchia:2019kta}.

The two-loop amplitudes for $\mathcal N \ge 4$ supergravity are
given in Ref.~\cite{BoucherVeronneau:2011qv}. The $\NeqEight$
supergravity result is the simplest of these and was
first given in Ref.~\cite{N8FiniteRemainder}
by combining the integrand of Ref.~\cite{BDDPR}
with the integrals of Ref.~\cite{DoubleBox}. Explicit results
for the finite remainders in the
Regge limit are found in Eqs.~(4.13)--(4.16) of
Ref.~\cite{Melville:2013qca}. 
 Note that the remainders in
Ref.~\cite{Melville:2013qca} are normalized with an extra factor of
$q^2 s$ relative to ours.

So far we have presented the classical scattering amplitudes in
perturbation theory, which assumes $Gs\ll1$. Ultimately, we are
interested in the limit $Gs\gg1$, with $Gs/J\ll1$ corresponding to the
classical post-Minkowskian expansion.  Implicitly this assumes that
the relevant parts of the perturbative series have been resummed. 
Standard ways to do so use eikonal or partial wave methods which
we utilize in the following.

\begin{figure}[t]
\includegraphics[scale=.47]{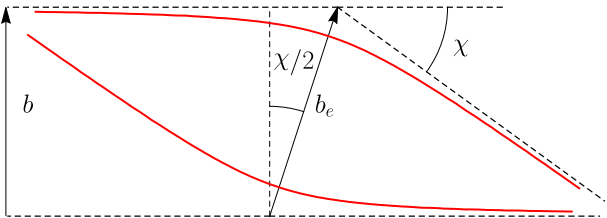}
\vskip -.3 cm
\caption{The scattering configuration showing the impact parameter, $b$, eikonal impact
  parameter, $b_e$, and the scattering angle, $\chi$. 
}
\vskip -.4 cm
\label{ScatteringFigure}
\end{figure}

\Section{Scattering angle from eikonal phase}
Following the usual procedure~\cite{EikonalReview, ACV, EikonalRecent,DiVecchia:2019kta},
we obtain the eikonal phase by taking the transverse Fourier transform of the
amplitude in the classical limit,
\begin{equation}
		\label{eikonal}
  -i \left( e^{i2\delta(s,b_e)} -1 \right) = \int \frac{\mu^{2\epsilon} d^{2-2\epsilon}q}{(2\pi)^{2-2\epsilon}} \, e^{i\vec q\cdot \vec b_e}\, \frac{\mathcal{M}(s,q^2)}{2s\mathcal{K}} \,,
\end{equation}
where $\delta(s, b_e)$ is the eikonal phase, which we expand perturbatively in Newton's constant 
$(\delta = \delta^{(0)} + \delta^{(1)} + \delta^{(2)} + \cdots)$, $\vec q$ is the $(2-2\epsilon)$-dimensional
vector in the scattering plane such that $\vec q^{\,2}=q^2$ and $b_e \equiv |\vec
b_e|$ is the eikonal impact parameter shown
in \fig{ScatteringFigure}. 
The basic formula needed for calculating the Fourier
transform is given in Eq.~(2.11) of Ref.~\cite{KoemansCollado:2019ggb}.

The full phase shift is generically complex, and be readily obtained from
Eqs.~\eqref{eq:TwoloopGenericAmpl}, \eqref{GRFiniteRemainder} and
\eqref{FiniteTwoLoop}. Its imaginary part at a given order captures inelastic (e.g., radiation)
effects. Here we are only interested in the conservative part, as in
Ref.~\cite{ACV} so we do not display it in the following and focus
only on the elastic phase. However, these imaginary parts are needed to extract the elastic
contributions at higher orders because of the exponentiation. The Fourier transform of polynomial terms
corresponds to short-range contact interactions, which are not relevant for the
problem of long-range scattering.

The universal $\mathcal{O}(G)$ result for the eikonal phase extracted
from the tree amplitude is
\begin{equation}
 \delta^{(0)} = \frac{Gs}{2} \left(\bar\mu^2 \tilde b_e^2\right)^{\epsilon} 
 \biggl[ -\frac{1}{\epsilon} - \frac12 \epsilon \zeta_2 
   -\frac13 \epsilon^2 \zeta_3 + \mathcal{O}(\epsilon^3) \biggr]\,,
\end{equation}
where we introduced $\tilde b_e = e^{\gamma_E}  b_e /2$ for
convenience.

As explained above, the pieces relevant for the one-loop scattering angle
are given by the real part of the nonanalytic part,
\begin{equation}
  \text{Re}\, F^{(1)}  = -\frac{\mathcal{N}-4}{2} L^2 
		+c L + \cdots \,,
\label{eq:F1nonuniv}
\end{equation}
where $\mathcal{N}$ denotes the amount of supersymmetry and $c$ is a
constant that takes on the values $0,-1, -87/10$ for $\mathcal{N}>4$,
$\mathcal{N}=4$ and pure gravity respectively.  The leading logarithms
($L^2$) arise from backward-scattering
diagrams~\cite{Melville:2013qca} and the subleading logarithm ($L$)
from bubble integrals. We conclude that they are nonuniversal and
depend on the specific theory. As mentioned above, the
$\Ord(G^2)$ one-loop phase can contribute to the angle only at the
quantum level, so this nonuniversality does not affect the classical
scattering angle. These contributions, including the $\Ord(\epsilon)$
parts, are however crucial for extracting the $\Ord(G^3)$ classical
pieces because of cross terms with infrared singularities.

The $\mathcal{O}(G^2)$ phase extracted from the one-loop amplitude is 
\begin{align}
\text{Re}\,  \delta^{(1)} = &  \frac{2G^2s}{\pi b_e^2}\left(\bar\mu^2 \tilde b_e^2\right)^{2\epsilon}  \label{SusyOneLoopPhase} \\ 
& \null \times  \biggl[ \frac{1}{\epsilon} -\frac{({\cal N} - 6)}{2}  \log(s\tilde b_e^2  ) 
+  \frac{c+2}{2} + \mathcal{O}(\epsilon) \biggr]\,, \nn
\end{align}
where $c$ is the same theory-dependent constant appearing in
Eq.~\eqref{eq:F1nonuniv}.  Additionally, there is an imaginary
part at $\mathcal{O}(\epsilon)$, needed to obtain the real
part of $\delta^{(2)}$, which is not displayed here but is readily
  obtained from the Fourier transform of the full amplitudes in
  \eqns{eq:TwoloopGenericAmpl}{GRFiniteRemainder} as well from
  Refs.~\cite{DunbarNorridge,InternalNotesGravity,TwoLoopGravityUV}.



The relevant terms at two loops arise from the nonanalytic terms in the
imaginary part of the remainder at one loop and from the real part at two
loops 
\begin{align} 
\text{Im}\,F^{(1)} &=  2 \pi L  - \epsilon \pi L^2 + \cdots  \,, \nn \\ 
\text{Re}\,F^{(2)}  &=  -2 \pi^2 L^2 + 4 \pi^2 L + \cdots  \,.
\label{ImportantTerms}
\end{align}
where the dots indicate non-universal terms which do not contribute to the phase at $\mathcal{O}(\epsilon^0)$.  This includes non-universal $\epsilon L$ terms in $\text{Im}\,F^{(1)}$ that could naively contribute but ultimately cancels against the iteration $- 2i \delta^{(0)}\delta^{(1)}$ coming from expanding the exponential.

The $\mathcal{O}(G^3)$ terms in the phase can thus be extracted from 
the two-loop amplitude after subtracting the iteration from the leading and subleading phases in the exponential \eqref{eikonal}.
The leading eikonal exponentiation also predicts a universal $\mathcal{O}(\epsilon)$ contribution to the two-loop amplitude which needs to be taken into account. (See the discussion in Ref.~\cite{DiVecchia:2019kta} near Eq.~(3.7)). We obtain the universal result,
\begin{equation}
\text{Re}\,  \delta^{(2)} = \frac{2G^3s^2}{b_e^2} \left(\bar\mu^2 \tilde b_e^2\right)^{3\epsilon} 
 + \mathcal{O}(\epsilon) \,,
\label{Eikonal2}
\end{equation}
valid for $\mathcal N \ge 4$ supergravity as well as pure Einstein
gravity. We are not displaying the imaginary parts since they are not universal and do not
contribute to the conservative dynamics at this order.

The classical scattering angle is given in terms of the eikonal phase via the
usual stationary-phase argument (see e.g. \cite{HighEnergyQuantum}),
\begin{equation}
  \sin \frac{1}{2} \chi(s,b_e) = 
- \frac{2}{\sqrt{s}} \frac{\partial}{\partial b_e}\, \delta(s,b_e) \,.
\end{equation}
Applying this formula to \eqn{Eikonal2}, which holds for all theories
evaluated here, we obtain the universal result
\begin{equation}
\sin\frac{1}{2} \chi(s,b_e) 
= \frac{2G\sqrt{s}}{b_e} +  \frac{(2G\sqrt{s})^3}{b_e^3}\,,
\label{eq:anglebe}
\end{equation}
matching the ACV pure gravity angle given in Eq.~(5.28) in
Ref.~\cite{ACV}, as well as the recently obtained angle in $\NeqEight$
supergravity~\cite{DiVecchia:2019kta}.
The scheme dependence cancels, as expected.  The result above is
written in terms of the symmetric impact parameter, $\vec b_e$ which
appears naturally in the eikonal formula.  This points in the
direction of the momentum transfer $\vec q$, while the more familiar impact
parameter $\vec b$ is perpendicular to the incoming momenta, as shown in
\fig{ScatteringFigure}. (See also Ref.~\cite{Ciafaloni:2014esa}.)   The relation between their
magnitudes is $b=b_e \cos(\chi/2)$. Rewriting the universal scattering
angle in terms of the usual impact parameter $b$ gives,
\begin{equation}
\sin\frac{1}{2} \chi(s,b) 
= \frac{2G\sqrt{s}}{b} + \frac{1}{2} \frac{(2G\sqrt{s})^3}{b^3}\,. 
\label{eq:angleb}
\end{equation}
We note that the quantum corrections to the scattering angle do not
display a corresponding universality, analogous to previously
observed nonuniversal spin dependence in quantum
corrections~\cite{QuantumNonUniversality}.

\Section{Scattering angle from partial-wave expansion}
Alternatively, we can extract the scattering angle from the
partial-wave expansion of the amplitude (see
e.g. Ref.~\cite{DamourRecent}).  Here we note that the partial waves
are given by
\begin{align}
  &a_l(s) =\frac{(16\pi\mu^2/s)^{\epsilon}}{ \Gamma(1-\epsilon)}  
  \int_{-1}^1 dx\, (1-x^2)^{-\epsilon}\, C_l^{\frac{1-2\epsilon}{2}}(x)\, \frac{\mathcal{M}(s,x)}{16\pi\, \mathcal K } \,, 
\end{align}
where $x=\cos\chi = 1+2t/s$ and the $C_l^{\frac{1-2\epsilon}{2}}(x)$ are  Geigenbauer polynomials (normalized to take unit value at $x=1$), which reduce to the more familiar Legendre polynomials when $\epsilon\rightarrow0$. 

If we ignore inelastic contributions, the partial waves can be parametrized in terms of phase shifts as
\begin{equation}
  a_l(s) =  -i \left( e^{i2\delta_l(s)} -1 \right)\,,
\end{equation}
and once again a stationary-phase argument gives the scattering angle as
\begin{equation}
  \frac12 \chi(s,l) =  -\frac{\partial\delta_l(s)}{\partial l}\,.
\end{equation}
Using this approach we find the phase shifts,
\begin{align}
  &\delta^{(0)}_l(s) =   \frac{Gs}{2}\left(\frac{\bar\mu^2 \tilde J^2}{s}\right)^{\epsilon} \!\left[ -\frac{1}{\epsilon} - \frac{1}{3J^2} + \mathcal{O}(\epsilon,J^{-4}) \right]  \,, \nn\\
  &\text{Re}\, \delta^{(1)}_l(s) =   \frac{ G^2s^2}{2\pi J^2}\left(\frac{\bar\mu^2 \tilde J^2}{s}\right)^{2\epsilon} \nn
  \\ &\hspace{0.8cm} \times  \left[ \frac{1}{\epsilon} -\frac{({\cal N} - 6)}{2}  \log(\tilde J^2  ) +  \frac{c+2}{2} + \mathcal{O}(\epsilon,J^{-2}) \right] \,, \nn \\
  &\text{Re}\, \delta^{(2)}_l(s) =  \frac{G^3s^3}{3J^2}  \left(\frac{\bar\mu^2 \tilde J^2}{s}\right)^{3\epsilon}  + \mathcal{O}(\epsilon,J^{-4})\,,
\end{align}
where $\tilde J^2 = e^{2\gamma_E}J^2$ and $J^2$ denotes the Casimir of the rotation group, i.e., $J^2:\equiv l(l+1-2\epsilon)$, which has a well defined classical limit.
The classical deflection angle is then
\vspace{-0.3cm}
\begin{equation}
  \frac12 \chi(s,J) = \frac{Gs}{J} + \frac{2}{3}\frac{G^3s^3}{J^3}\,,
\end{equation}
written in terms of the classical variables, or, equivalently,
\vspace{-0.2cm}
\begin{equation}
  \sin\frac12 \chi(s,J) = \frac{Gs}{J} +\frac{1}{2} \frac{G^3s^3}{J^3}\,.
  \label{eq:angleJ}
\end{equation}
Using the relation between the angular momentum and the impact parameters
\begin{equation}
  J = \frac{\sqrt{s}}{2} b = \frac{\sqrt{s}}{2} b_e \cos \frac12 \chi\,,
\end{equation}
we find that \eqn{eq:angleJ} reproduces \eqns{eq:anglebe}{eq:angleb}.

We can directly compare our results to Damour's angle given in
Eq.~(5.37) of Ref.~\cite{DamourRecent}, following from a conjecture of smooth
high-energy behavior between the massless and massive cases,
\begin{equation}
  \sin \frac{1}{2} \chi^{\rm D}(s,J) = \frac{G s}{J} - \frac{3}{4} \frac{G^3 s^3}{J^3} \,.
\label{DamourAngle}
\end{equation} 
As noted in Ref.~\cite{DamourRecent}, this disagrees with the angle
obtained by ACV, which is matched by \eqn{eq:angleJ}. As
emphasized by Damour~\cite{DamourRecent}, because the sign of the
$G^3$ term in \eqn{DamourAngle} is opposite to that of 
\eqn{eq:angleJ} the disagreement between the two formulas is robust.

Here we focused on the scattering of identical-helicity gravitons in
the initial state.  We have repeated the calculation for the case of
opposite-helicity gravitons with the same results for the classical
scattering angle.  Furthermore, we expect the result to be identical
for any massless external states.  Indeed, for the supersymmetric
cases that we analyzed, supersymmetry identities~\cite{SWI} relate
graviton scattering to scattering of other massless states.

\Section{Conclusions}
By studying gravitational scattering amplitudes through $\Ord(G^3)$ in
a variety of theories, we found the classical scattering angle to be
independent of their matter content, thus demonstrating graviton
dominance at a higher order than had been previously
understood~\cite{HighEnergyQuantum}.  In addition, we confirmed that
the classical scattering angle found by ACV~\cite{ACV} is indeed
correct.  The results of our calculation are, however, in conflict
with Damour's recent conjecture~\cite{DamourRecent}.

There are a number of interesting directions to pursue.  First and
foremost, it would be desirable to systematically complete a proof of
universality through $\Ord(G^3)$ for any massless gravitational
theory. An obvious, if nontrivial, next step would be to check whether
some form of universality remains at higher orders as well.  It would
also be important to understand the constraints that the high-energy
behavior of scattering amplitudes imposes on classical binary black
hole interactions~\cite{DamourHighEnergy}.  The recent
advances~\cite{TwoLoopNumericalUnitarity,TwoLoopNumericalUnitarityMultiscale}
that make it possible to obtain the complete four-graviton two-loop
amplitude of pure Einstein gravity~\cite{FreiburgN0Ampl} can be
expected to lead to further advances, including for the important
case of massive multiloop amplitudes relevant for the gravitational-wave 
two-body problem.

\Section{Acknowledgments}
We thank Alessandra Buonanno, Clifford Cheung, Marcello Ciafaloni, Thibault
Damour, Paolo Di Vecchia, David Kosower, Andr\'es Luna, Stephen Naculich, Radu
Roiban, Rodolfo Russo, Donal O'Connell, Chia-Hsien Shen, Mikhail Solon, Jan
Steinhoff, Justin Vines, Zahra Zahraee and especially Simon Caron-Huot and
Gabriele Veneziano for many helpful comments and discussions.  We thank the
U.S. Department of Energy (DOE) for support under grant no.~DE-SC0009937. JPM
is supported by the US Department of State through a Fulbright scholarship.
M.S.R.'s work is funded by the German Research Foundation (DFG) within the
Research Training Group GRK 2044.  We are also grateful to the Mani L. Bhaumik
Institute for Theoretical Physics for generous support.



\begin{thebibliography}{99}

\bibitem{HighEnergyClassical} 
P.~D.~D'Eath,
Phys.\ Rev.\ D {\bf 18}, 990 (1978);
%
S.~J.~Kovacs and K.~S.~Thorne,
Astrophys.\ J.\  {\bf 217}, 252 (1977);
%
S.~J.~Kovacs and K.~S.~Thorne,
Astrophys.\ J.\  {\bf 224}, 62 (1978).

\bibitem{HighEnergyQuantum}
D.~Amati, M.~Ciafaloni and G.~Veneziano,
Phys.\ Lett.\ B {\bf 197}, 81 (1987);
%
G.~'t Hooft,
Phys.\ Lett.\ B {\bf 198}, 61 (1987);
%
D.~Amati, M.~Ciafaloni and G.~Veneziano,
Int.\ J.\ Mod.\ Phys.\ A {\bf 3}, 1615 (1988);
%
I.~J.~Muzinich and M.~Soldate,
Phys.\ Rev.\ D {\bf 37}, 359 (1988).


\bibitem{ACV} 
D.~Amati, M.~Ciafaloni and G.~Veneziano,
Nucl.\ Phys.\ B {\bf 347}, 550 (1990).

\bibitem{LIGO}
B.~P.~Abbott {\it et al.} [LIGO Scientific and Virgo Collaborations],
  Phys.\ Rev.\ Lett.\  {\bf 116}, no. 6, 061102 (2016)
  [arXiv:1602.03837 [gr-qc]];
%
 B.~P.~Abbott {\it et al.} [LIGO Scientific and Virgo Collaborations],
  Phys.\ Rev.\ Lett.\  {\bf 119}, no. 16, 161101 (2017)
  [arXiv:1710.05832 [gr-qc]].

\bibitem{ScatteringToPotential}
Y.~Iwasaki,
Prog.\ Theor.\ Phys.\  {\bf 46}, 1587 (1971);
%
Y.~Iwasaki,
Lett.\ Nuovo Cim.\  {\bf 1S2}, 783 (1971)
[Lett.\ Nuovo Cim.\  {\bf 1}, 783 (1971)];
%
H.~Okamura, T.~Ohta, T.~Kimura and K.~Hiida,
Prog.\ Theor.\ Phys.\  {\bf 50}, 2066 (1973);
%
S.~N.~Gupta and S.~F.~Radford,
Phys.\ Rev.\ D {\bf 19}, 1065 (1979);
%
J.~F.~Donoghue,
  Phys.\ Rev.\ D {\bf 50}, 3874 (1994)
  [gr-qc/9405057];
%
B.~R.~Holstein and J.~F.~Donoghue,
  Phys.\ Rev.\ Lett.\  {\bf 93}, 201602 (2004)
  [hep-th/0405239];
%
D.~Neill and I.~Z.~Rothstein,
Nucl.\ Phys.\ B {\bf 877}, 177 (2013)
[arXiv:1304.7263 [hep-th]];
%
V.~Vaidya,
Phys.\ Rev.\ D {\bf 91}, no. 2, 024017 (2015)
[arXiv:1410.5348 [hep-th]];
%
N.~E.~J.~Bjerrum-Bohr, J.~F.~Donoghue and P.~Vanhove,
JHEP {\bf 1402}, 111 (2014)
[arXiv:1309.0804 [hep-th]];
%
N.~E.~J.~Bjerrum-Bohr, P.~H.~Damgaard, G.~Festuccia, L.~Plant\'e and P.~Vanhove,
Phys.\ Rev.\ Lett.\  {\bf 121}, no. 17, 171601 (2018)
[arXiv:1806.04920 [hep-th]];
%
C.~Cheung, I.~Z.~Rothstein and M.~P.~Solon,
Phys.\ Rev.\ Lett.\  {\bf 121}, no. 25, 251101 (2018)
[arXiv:1808.02489 [hep-th]];
%
D.~A.~Kosower, B.~Maybee and D.~O'Connell,
JHEP {\bf 1902}, 137 (2019)
[arXiv:1811.10950 [hep-th]].

\bibitem{3PMPRL} 
Z.~Bern, C.~Cheung, R.~Roiban, C.~H.~Shen, M.~P.~Solon and M.~Zeng,
Phys.\ Rev.\ Lett.\  {\bf 122}, no. 20, 201603 (2019)
[arXiv:1901.04424 [hep-th]].

\bibitem{3PMLong} 
Z.~Bern, C.~Cheung, R.~Roiban, C.~H.~Shen, M.~P.~Solon and M.~Zeng,
JHEP {\bf 1910}, 206 (2019)
[arXiv:1908.01493 [hep-th]].

\bibitem{GeneralizedUnitarity}
Z.~Bern, L.~J.~Dixon, D.~C.~Dunbar and D.~A.~Kosower,
Nucl.\ Phys.\ B {\bf 425}, 217 (1994)
[hep-ph/9403226];
%
Z.~Bern, L.~J.~Dixon, D.~C.~Dunbar and D.~A.~Kosower,
Nucl.\ Phys.\ B {\bf 435}, 59 (1995)
[hep-ph/9409265];
%
Z.~Bern, L.~J.~Dixon and D.~A.~Kosower,
Nucl.\ Phys.\ B {\bf 513}, 3 (1998)
[hep-ph/9708239];
%
R.~Britto, F.~Cachazo and B.~Feng,
Nucl.\ Phys.\ B {\bf 725}, 275 (2005)
[hep-th/0412103];
%
Z.~Bern, J.~J.~M.~Carrasco, H.~Johansson and D.~A.~Kosower,
Phys.\ Rev.\ D {\bf 76}, 125020 (2007)
[arXiv:0705.1864 [hep-th]].


\bibitem{DoubleCopy}
H.~Kawai, D.~C.~Lewellen and S.~H.~H.~Tye,
Nucl.\ Phys.\ B {\bf 269}, 1 (1986);
%
Z.~Bern, L.~J.~Dixon, M.~Perelstein and J.~S.~Rozowsky,
Nucl.\ Phys.\ B {\bf 546}, 423 (1999)
[hep-th/9811140];
%
Z.~Bern, J.~J.~M.~Carrasco and H.~Johansson,
Phys.\ Rev.\ D {\bf 78}, 085011 (2008)
[arXiv:0805.3993 [hep-ph]]l;
%
Z.~Bern, J.~J.~M.~Carrasco and H.~Johansson,
Phys.\ Rev.\ Lett.\  {\bf 105}, 061602 (2010)
[arXiv:1004.0476 [hep-th]];
%
Z.~Bern, J.~J.~Carrasco, M.~Chiodaroli, H.~Johansson and R.~Roiban,
arXiv:1909.01358 [hep-th].

\bibitem{DamourHighEnergy} 
T.~Damour,
Phys.\ Rev.\ D {\bf 97}, no. 4, 044038 (2018)
[arXiv:1710.10599 [gr-qc]].

\bibitem{EOB}
A.~Buonanno and T.~Damour,
Phys.\ Rev.\ D {\bf 59}, 084006 (1999)
[gr-qc/9811091];
%
A.~Buonanno and T.~Damour,
Phys.\ Rev.\ D {\bf 62}, 064015 (2000)
[gr-qc/0001013].

\bibitem{DamourRecent} 
T.~Damour,
arXiv:1912.02139 [gr-qc].

\bibitem{SelfForce}
Y.~Mino, M.~Sasaki and T.~Tanaka,
  Phys.\ Rev.\ D {\bf 55}, 3457 (1997)
  [gr-qc/9606018];
%
T.~C.~Quinn and R.~M.~Wald,
particles in curved space-time,''
Phys.\ Rev.\ D {\bf 56}, 3381 (1997)
[gr-qc/9610053].

\bibitem{Bellini:1992eb} 
  A.~Bellini, M.~Ademollo and M.~Ciafaloni,
  Nucl.\ Phys.\ B {\bf 393}, 79 (1993)
  [hep-th/9207113].


\bibitem{BDDPR}
Z.~Bern, L.~J.~Dixon, D.~C.~Dunbar, M.~Perelstein and J.~S.~Rozowsky,
Nucl.\ Phys.\ B {\bf 530}, 401 (1998)
[hep-th/9802162].

\bibitem{N8FiniteRemainder}
S.~G.~Naculich, H.~Nastase and H.~J.~Schnitzer,
Nucl.\ Phys.\ B {\bf 805}, 40 (2008)
[arXiv:0805.2347 [hep-th]].

\bibitem{BoucherVeronneau:2011qv}
C.~Boucher-Veronneau and L.~J.~Dixon,
JHEP {\bf 1112}, 046 (2011)
[arXiv:1110.1132 [hep-th]].

\bibitem{FreiburgN0Ampl} 
S.~Abreu, M.~Jaquier, F.~Febres Cordero, H.~Ita, B.~Page, 
M.~S.~Ruf, and V.~Sotnikov, 
to appear.

\bibitem{TwoLoopNumericalUnitarity}
H.~Ita,
Phys.\ Rev.\ D {\bf 94}, no. 11, 116015 (2016)
[arXiv:1510.05626 [hep-th]].
%
S.~Abreu, F.~Febres Cordero, H.~Ita, M.~Jaquier, B.~Page and M.~Zeng,
Phys.\ Rev.\ Lett.\  {\bf 119}, no. 14, 142001 (2017)
[arXiv:1703.05273 [hep-ph]].

\bibitem{EikonalReview}
R.~J.~Glauber,
Lectures in Theoretical Physics, ed.~by W.~E. Brittin and L.~G.~Dunham,
Interscience Publishers, Inc., New York, Volume I, page 315, (1959).

\bibitem{EikonalRecent}
E.~Laenen, G.~Stavenga and C.~D.~White,
JHEP {\bf 0903}, 054 (2009)
[arXiv:0811.2067 [hep-ph]];
%
S.~B.~Giddings, M.~Schmidt-Sommerfeld and J.~R.~Andersen,
Phys.\ Rev.\ D {\bf 82}, 104022 (2010)
[arXiv:1005.5408 [hep-th]];
%
P.~Di Vecchia, A.~Luna, S.~G.~Naculich, R.~Russo, G.~Veneziano and C.~D.~White,
Phys.\ Lett.\ B {\bf 798}, 134927 (2019)
[arXiv:1908.05603 [hep-th]];
%
R.~Akhoury, R.~Saotome and G.~Sterman,
arXiv:1308.5204 [hep-th];
%
M.~Kulaxizi, G.~S.~Ng and A.~Parnachev,
JHEP {\bf 1910}, 107 (2019)
[arXiv:1907.00867 [hep-th]];

\bibitem{Melville:2013qca} 
S.~Melville, S.~G.~Naculich, H.~J.~Schnitzer and C.~D.~White,
Phys.\ Rev.\ D {\bf 89}, no. 2, 025009 (2014)
[arXiv:1306.6019 [hep-th]].

\bibitem{Ciafaloni:2014esa} 
  M.~Ciafaloni and D.~Colferai,
  JHEP {\bf 1410}, 085 (2014)
  [arXiv:1406.6540 [hep-th]].


\bibitem{KoemansCollado:2019ggb} 
  A.~Koemans Collado, P.~Di Vecchia and R.~Russo,
  Phys.\ Rev.\ D {\bf 100}, no. 6, 066028 (2019)
  [arXiv:1904.02667 [hep-th]].

\bibitem{DiVecchia:2019kta} 
P.~Di Vecchia, S.~G.~Naculich, R.~Russo, G.~Veneziano and C.~D.~White,
arXiv:1911.11716 [hep-th].

\bibitem{HennThreeLoopN8}
J.~M.~Henn and B.~Mistlberger,
JHEP {\bf 1905}, 023 (2019)
[arXiv:1902.07221 [hep-th]].

\bibitem{GravityIR}
S.~Weinberg,
Phys.\ Rev.\  {\bf 140}, B516 (1965)
%
S.~G.~Naculich and H.~J.~Schnitzer,
JHEP {\bf 1105}, 087 (2011)
[arXiv:1101.1524 [hep-th]].

\bibitem{DunbarNorridge}
D.~C.~Dunbar and P.~S.~Norridge,
Nucl.\ Phys.\ B {\bf 433}, 181 (1995)
[hep-th/9408014].

\bibitem{InternalNotesGravity}
Z.~Bern, C.~Cheung, H.-H.~Chi, S.~Davies, L.~Dixon, and
J.~Nohle, unpublished.

\bibitem{TwoLoopGravityUV}
Z.~Bern, C.~Cheung, H.~H.~Chi, S.~Davies, L.~Dixon and J.~Nohle,
Phys.\ Rev.\ Lett.\  {\bf 115}, no. 21, 211301 (2015)
[arXiv:1507.06118 [hep-th]];
%
Z.~Bern, H.~H.~Chi, L.~Dixon and A.~Edison,
Phys.\ Rev.\ D {\bf 95}, no. 4, 046013 (2017)
[arXiv:1701.02422 [hep-th]].

\bibitem{Bern:2011rj} 
Z.~Bern, C.~Boucher-Veronneau and H.~Johansson,
Phys.\ Rev.\ D {\bf 84}, 105035 (2011)
[arXiv:1107.1935 [hep-th]].

\bibitem{tHooftVeltman} 
G.~'t Hooft and M.~J.~G.~Veltman,
Ann.\ Inst.\ H.\ Poincare Phys.\ Theor.\ A {\bf 20}, 69 (1974).

\bibitem{GoroffSagnotti} 
M.~H.~Goroff and A.~Sagnotti,
Nucl.\ Phys.\ B {\bf 266}, 709 (1986);
A.~E.~M.~van de Ven,
Nucl.\ Phys.\ B {\bf 378}, 309 (1992).

\bibitem{DoubleBox}
V.~A.~Smirnov,
Phys.\ Lett.\ B {\bf 460}, 397 (1999)
[hep-ph/9905323];
%
J.~B.~Tausk,
Phys.\ Lett.\ B {\bf 469}, 225 (1999)
[hep-ph/9909506].


\bibitem{QuantumNonUniversality} 
N.~E.~J.~Bjerrum-Bohr, J.~F.~Donoghue, B.~R.~Holstein, L.~Planté and P.~Vanhove,
Phys.\ Rev.\ Lett.\  {\bf 114}, no. 6, 061301 (2015)
[arXiv:1410.7590 [hep-th]];
%
N.~E.~J.~Bjerrum-Bohr, J.~F.~Donoghue, B.~R.~Holstein, L.~Plante and P.~Vanhove,
JHEP {\bf 1611}, 117 (2016)
[arXiv:1609.07477 [hep-th]];
%
D.~Bai and Y.~Huang,
Phys.\ Rev.\ D {\bf 95}, no. 6, 064045 (2017)
[arXiv:1612.07629 [hep-th]];
%
H.~H.~Chi,
Phys.\ Rev.\ D {\bf 99}, no. 12, 126008 (2019)
[arXiv:1903.07944 [hep-th]].

\bibitem{SWI}
M.~T.~Grisaru, H.~N.~Pendleton and P.~van Nieuwenhuizen,
Phys.\ Rev.\ D {\bf 15}, 996 (1977);
%
M.~T.~Grisaru and H.~N.~Pendleton,
Nucl.\ Phys.\ B {\bf 124}, 81 (1977).

\bibitem{TwoLoopNumericalUnitarityMultiscale}
S.~Abreu, F.~Febres Cordero, H.~Ita, B.~Page and M.~Zeng,
Phys.\ Rev.\ D {\bf 97}, no. 11, 116014 (2018)
[arXiv:1712.03946 [hep-ph]];
%
S.~Abreu, J.~Dormans, F.~Febres Cordero, H.~Ita and B.~Page,
Phys.\ Rev.\ Lett.\  {\bf 122}, no. 8, 082002 (2019)
[arXiv:1812.04586 [hep-ph]].


\end{thebibliography}
\end{document}